\documentclass[utf8]{revtex4-2}

\usepackage{graphicx}
\usepackage{multirow}
\usepackage{array}
\usepackage{amssymb}
\newcolumntype{P}[1]{>{\centering\arraybackslash}p{#1}}
\newcolumntype{M}[1]{>{\centering\arraybackslash}m{#1}}

\usepackage{hyperref}
\usepackage[normalem]{ulem}

\usepackage{placeins}

\usepackage{soul}

\usepackage{color}

\begin{document}

\title{Superconducting spin valve effect in the  Co/Pb/Co heterostructure\\  with insulating interlayers}

\author{A.~A.~Kamashev}
\affiliation{Zavoisky Physical-Technical Institute, FRC Kazan
Scientific Center of RAS, 420029 Kazan, Russia}

\author{N.~N.~Garif'yanov}
\affiliation{Zavoisky Physical-Technical Institute, FRC Kazan
Scientific Center of RAS, 420029 Kazan, Russia}

\author{A.~A.~Validov}
\affiliation{Zavoisky Physical-Technical Institute, FRC Kazan
Scientific Center of RAS, 420029 Kazan, Russia}

\author{V.~Kataev}
\affiliation{Leibniz Institute for Solid State and Materials
	Research, Helmholtzstr. 20, D-01069 Dresden, Germany}

\author{A.~S.~Osin}
\affiliation{L.~D.\ Landau Institute for Theoretical Physics RAS, 142432 Chernogolovka, Russia}

\author{Ya.~V.~Fominov}
\affiliation{L.~D.\ Landau Institute for Theoretical Physics RAS, 142432 Chernogolovka, Russia}
\affiliation{Laboratory for Condensed Matter Physics, HSE University, 101000 Moscow, Russia}

\author{I.~A.~Garifullin}
\affiliation{Zavoisky Physical-Technical Institute, FRC Kazan Scientific Center of RAS, 420029 Kazan, Russia}

\date{\today}

\begin{abstract}
We report the superconducting properties of the  Co/Pb/Co heterostructures with thin insulating interlayers. The main specific feature
of these structures is the intentional oxidation of both superconductor/ferromagnet (S/F) interfaces.
We study variation of the critical temperature of our systems due to switching between parallel and antiparallel configurations of the magnetizations of the two magnetic layers.
Common wisdom suggests that this spin valve effect, which is due to the S/F proximity effect, is most pronounced in the case of perfect metallic contact at the interfaces.
Nevertheless, in our structures with intentionally deteriorated interfaces, we observed a significant full spin valve effect.
A shift of the superconducting transition temperature $T_c$ by switching the mutual orientation of the magnetizations of the two ferromagnetic Co layers from antiparallel to parallel amounted to  $\Delta T_c =0.2$\,K at the optimal thickness of the superconducting Pb layer. Our finding verifies the so far unconfirmed earlier results by Deutscher and Meunier on an F1/S/F2 heterostructure with oxidized interlayers  [\href{https://doi.org/10.1103/PhysRevLett.22.395}{G.\ Deutscher and F.\ Meunier, Phys.\ Rev.\ Lett.\ \textbf{22}, 395 (1969)}] and suggests an alternative route to optimize the performance of the superconducting spin valves.
\end{abstract}

\maketitle


\section{Introduction}

Models and specific realizations of the superconducting spin valve (SSV) have been the subject of intensive research  over the past 25 years
(see, e.g., \cite{Oh,Tagirov,Buzdin2,Gu,Moraru,Potenza,Westerholt,Steiner,Pugach2009,Leksin2010}). The interest in these structures is due to the possibility to observe and exploit the reciprocal influence of superconductivity (S) and ferromagnetism (F) on each other's properties when they are put into a close contact (see, e.g., \cite{Ioffe,Feigelman,Buzdin1,Bergeret,Blamire,Linder,Eschrig}). Moreover, the SSV structures appear as promising devices for applications in modern superconducting spintronics (see, e.g., \cite{Demler1997,Garifullin_obzor_2002,Zutic,Linder2009,Efetov_Springer_2013}). In 1997, Beasley and coworkers in proposed a theoretical F1/F2/S model of the SSV structure \cite{Oh}.
Another F1/S/F2 model was developed a little later in 1999 by Tagirov \cite{Tagirov} and Buzdin {\it et al.} \cite{Buzdin2}.
In these structures F1 and F2 are metallic ferromagnetic layers, and S is a superconducting layer.
Both models analyze the penetration of Cooper pairs from the S layer into the F layers under the action of the exchange field generated by F1 and F2 layers. Those early theoretical works implied operation principle of the SSV structure based on the control of the average exchange field acting on the S layer by changing the mutual orientation of the magnetization vectors of the F layers and thus  suppressing superconductivity to a different degree. Typically, the superconducting transition temperature $T_c$ of the SSV is minimal~/~maximal for the parallel (P)~/~antiparallel (AP) geometry of the two vectors, respectively.
The magnitude of the SSV effect is defined as the difference of these two temperatures  $\Delta T_c=T_c^\mathrm{AP}-T_c^\mathrm{P}$.
The full SSV effect is realized when $\Delta T_c$ is larger than the superconducting
transition width $\delta T_c$ in the P and AP configurations. Several experimental works confirmed the predicted influence of the mutual orientation of the magnetization vectors of the F layers
 on $T_c$ in the F1/S/F2 type of structures (see, e.g., \cite{Gu,Moraru,Potenza,You,Pena,Miao}). However,
a full switching between the normal and the superconducting state was not achieved because in these SSVs $\Delta T_c$ was always smaller than $\delta T_c$.
For the first time, a complete switching between  the normal and superconducting states was observed in the F1/F2/S type of SSV in Ref.\ \cite{Leksin2010}.

Theories (see, e.g., \cite{Buzdin1,Blamire,Linder,Bergeret2,Eschrig2011,Efetov_Springer_2007}) predict that at certain conditions a long-range triplet component (LRTC)
in the superconducting condensate can arise in the S/F bilayer. The generation of the LRTC opens an additional channel for the leakage of the Cooper pairs from the S into the  F layers in the F1/F2/S SSV at noncollinear configuration of F1 and F2 magnetizations, which significantly suppresses $T_c$ and thus should manifest as a minimum of
$T_c$ at the orthogonal magnetizations' geometry \cite{Fominov}. A large number of theoretical and experimental works have been devoted to
the study of this effect (see, e.g., \cite{Fominov,Leksin2012,Wu,Banerjee2,Leksin2015,Garifullin,Gu2015,Aarts2015,Leksin2016,Kamashev2019,Kamashev20191}).

By now, many such
SSVs using various elemental metals and alloys have been studied in sufficient detail and recent results indicate that
significant values
of the SSV effect have already been
achieved in the F1/F2/S structures (see, e.g., \cite{Gu2015,Aarts2015,Kamashev20191}).
Since the principle of an SSV relies on the S/F proximity effect which is confined to the interface between the S and F layers, a particular attention was given to the quality of this interface in terms of its morphology, smoothness, absence of intergrowth, etc. \cite{Nano},
which defines the mainstream approach in this field.
At odds with this approach, a significant SSV effect of $\Delta T_c \sim 0.3$\,K in the FeNi/In/Ni heterostructure with intentionally oxidized F/S interfaces
was demonstrated by Deutscher and Meunier in 1969 \cite{Deutscher}. The idea behind the oxidation of the FeNi and Ni layers was to slightly weaken the S/F proximity effect such that the superconductivity in the In layer could not be completely destroyed by the exchange field of the F layers. The authors noted that the thin oxidized layers became insulating but presumably remained magnetic. In a later experiment by Li {\it et al.} \cite{Li2013}, the F layers themselves were insulating by design. In this special situation,
even a very thin additional nonmagnetic insulating interlayer at the interface immediately suppressed the S/F proximity effect.

The paradoxical fact that ``worsening'' of the S/F interface in a metallic system \cite{Deutscher} can yield significant magnitude of the SSV effect is remarkable.

The work by Deutscher and Meunier~ \cite{Deutscher} has never been reproduced and the research in this direction was not pursued albeit, according to private communications in the SSV community, some groups attempted but did not succeed to reproduce that early remarkable result.

In the present work, in order to verify the SSV effect reported by Deutscher and Meunier for the heterostructure where the superconducting layer was

contacted to the ferromagnets through

thin insulating interlayers and to prove the validity of this concept to other types of the SSV structures, we investigated the superconducting properties of the SSV made of completely different F and S materials as compared to those in Ref.\ \cite{Deutscher}. Specifically, we prepared the multilayers Co1/Pb/Co2 with oxidized Co1/Pb and Pb/Co2 interfaces following the recipe of Refs.\ \cite{Lommel1968,Deutscher}.
We studied the dependence of the magnitude of the SSV effect $\Delta T_c$ on the Pb layer thickness and found that $\Delta T_c$ reached 0.2\,K for the optimal thickness, surpassing most of the values previously observed for the SSVs with perfect metallic contact. We discuss the obtained results in the context of the existing theoretical models of the S/F proximity effect.

\section{Samples}

For our investigation
CoO$_x$(3.5nm)/\allowbreak Co1(3nm)/\allowbreak I1/\allowbreak Pb($d_\mathrm{Pb}$)/\allowbreak I2/\allowbreak Co2(3nm)/\allowbreak Si$_3$N$_4$(85nm) heterostructures with the variable Pb-layer thickness $d_\mathrm{Pb}$ in the range from 60 to 120\,nm were fabricated on the high quality single crystalline substrate MgO (001).
Here, Co1 and Co2 are ferromagnetic F1 and F2 layers, I1 and I2 are thin oxide insulating interlayers, Pb is the superconducting layer, Si$_3$N$_4$
is a protective layer, and CoO$_x$ is the antiferromagnetic (AF) bias layer that fixes the direction of the magnetization
of the Co1 layer. The layers were deposited using an  electron beam evaporation  (Co, Pb) and AC sputtering (Si$_3$N$_4$) techniques.
The deposition setup had a load-lock station with vacuum shutters, allowing to transfer the sample holder without breaking the ultra-high vacuum in the deposition chambers.
The load-lock station provides possibility to oxidize the prepared layers in a controlled atmosphere. This allows to prepare the AF CoO$_x$ layer and fabricate the thin oxide interlayers (I1 and I2) at the Co1/I1/Pb and Pb/I2/Co2 interfaces. CoO$_x$ was prepared by exposing the metallic Co  layer to oxygen atmosphere at 100\,mbar during
two hours. Next, Co1 was deposited in the main deposition chamber at the vacuum of order $10^{-9}$\,mbar on top of the CoO$_x$ layer. The  I1 layer was formed on the surface of Co1 in the similar way as described above in the oxygen atmosphere of about  $\sim 10^{-2}$\,mbar during 60\,sec. It was shown in Ref.~\cite{Smardz} that significant partial oxidation of a few nanometers thin metallic Co layer can be achieved by exposing it to the ambient atmospheric environment implying that lowering the atmospheric pressure by five orders of magnitude enables one to oxidize only the surface while not affecting the bulk of the layer. After that, the Pb layer and subsequent layers of the SSV structure were deposited at the substrate temperature of $T_\mathrm{sub}\sim 150$\,K. Such low $T_\mathrm{sub}$ was necessary to obtain a smooth Pb layer \cite{Nano} and to form the I2 layer. The similar oxidation procedure was used again to form the I2 layer by exposing the Pb surface to oxygen atmosphere of about  $\sim 10^{-2}$\,mbar during 30\,sec. After that, the Co2 layer was deposited similar to the Co1 layer.

According to Refs.~ \cite{Lommel1968,Deutscher} the O$_2$ molecules adsorbed on the surface of the superconducting Pb layer oxidize the top ferromagnetic Co2 layer during its deposition thereby  forming an insulating magnetic interlayer at the S/F interface. We consider oxidation of the Pb layer to be unlikely because it was deposited at a low substrate temperature and exposed to a very low atmospheric pressure for a very short time, as specified above. According to the literature, the formation of an oxide on the surface of the Pb film requires significantly higher temperatures and pressures, and much longer exposition times (see, e.g., Refs.~\cite{Schroen1968,Hapase1968,Anderson1964}).

\begin{figure}
\center{\includegraphics[width=0.5\linewidth]{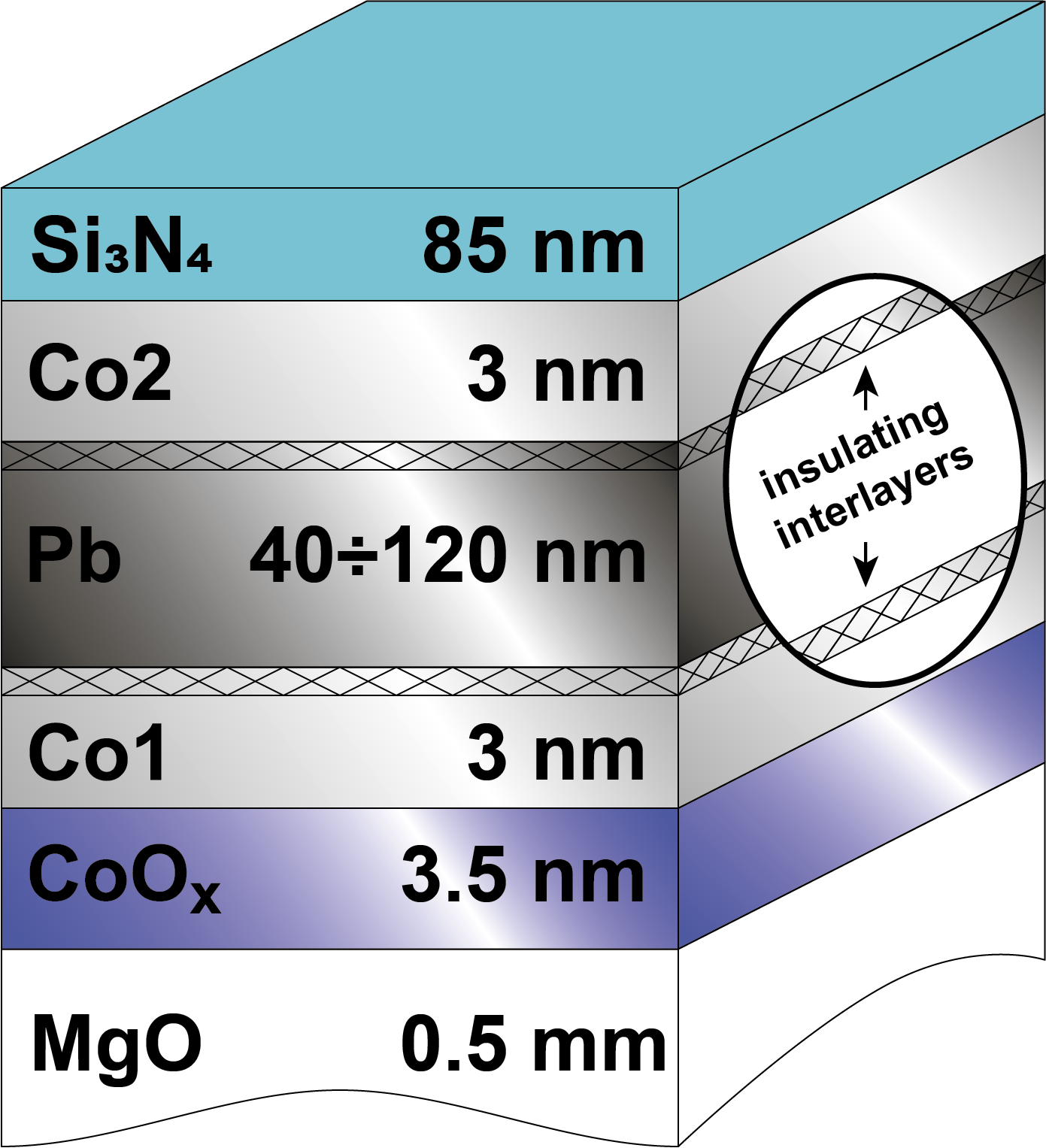}}
\caption{Design of the prepared samples. Cross-dashed areas depict I1 and I2 insulating interlayers (see the text for details).}
\label{fig1}
\end{figure}

Finally, all samples were covered with a protective Si$_3$N$_4$ layer.
The deposition rates were as follows: 0.5\,{\AA}/s for Co1 and Co2, 12\,{\AA}/s for  Pb, and 1.8\,{\AA}/s for Si$_3$N$_4$ films.
The final design of the samples is depicted in Fig.~\ref{fig1}.

Based on our previous studies in Refs.\ \cite{Leksin2015,Garifullin,JMMM} we chose the thickness of the AF CoO$_x$ layer to be $d_{\mathrm{CoO}_x} = 3.5$\,nm which is optimal to hold the direction of the magnetization of the Co1 layer up to the in-plane external magnetic field strength of $H_0^\mathrm{max} \sim 1.5$\,kOe. Magnetic studies of samples are presented in Supporting Information \ref{s1}.
We took the same thickness of 3\,nm for both Co1 and Co2 layers.

In addition, a control set of the samples with similar thicknesses of the S and F layers but without insulating interlayers at the Co1/Pb/Co2 interfaces was prepared for comparison.
The list of the studied samples CoO$_x$(3.5nm)/Co1(3nm)/I1/Pb($d_\mathrm{Pb}$)/I2/Co2(3nm)\-/Si$_3$N$_4$(85nm)
with the variable Pb layer thickness $d_\mathrm{Pb}$ is presented in Table~1.

\begin{table}
\caption{List of the studied samples CoO$_x$(3.5nm)/Co1(3nm)/I1/Pb($d_\mathrm{Pb}$)/I2/Co2(3nm)\-/Si$_3$N$_4$(85nm)
with the variable Pb layer thickness $d_\mathrm{Pb}$.}

\begin{tabular}{|P{2.5cm}|P{1.5cm}|P{3cm}|}

\hline	
{ Samples with insulating inter\-layers } & \raisebox{-4.5ex}{ $d_\mathrm{Pb}$ (nm)} \\ \hline
Pb\underline{ }120 & 120 \\ \hline
Pb\underline{ }100 & 100 \\ \hline
Pb\underline{ }80 & 80 \\  \hline
Pb\underline{ }60 & 60 \\ \hline
Pb\underline{ }40 & 40 \\ \hline
\end{tabular}
\end{table}

\section{Experimental results}
Electrical resistivity measurements were carried out with a standard four-point method in the DC mode.
For changing the mutual direction of the magnetization of the F layers between the P and AP orientations an external magnetic field of the order $\sim 1\,\mathrm{kOe} < H_0^\mathrm{max}$ was always applied in the plane of the sample in all measurements. The strength of the magnetic field was measured by a Hall probe with an accuracy of $\pm 0.3$\,Oe.
The sample temperature was monitored by the highly sensitive in the temperature range of interest carbon Allen–Bradley thermometer.
The temperature measurement error was $\pm\, (5-6)$\,mK below 3\,K.
The superconducting critical temperature $T_c$ was defined as the midpoint of the transition curve.

To study  the SSV effect the samples were cooled down from room to low temperatures in a magnetic field of the order 5\,kOe (field cooling procedure) applied in the sample plane. This field aligns magnetization of both F layers and the magnetization vector of the Co1 layer is getting fixed in the direction of the applied field and remains biased by the AF CoO$_x$ layer after the reduction of the field strength to its operational value of $H_0 = 1$\,kOe independent of the subsequent direction of the field vector \cite{Leksin2015,Garifullin,JMMM}. At this field value the temperature dependence of the resistivity $R(T)$ was recorded  for the P and AP configurations of the magnetizations of the Co1 and Co2 layers by appropriate rotation of the magnetization of the Co2 layer by the external magnetic field.

\begin{figure}
\begin{center}
\includegraphics[width=0.5\columnwidth]{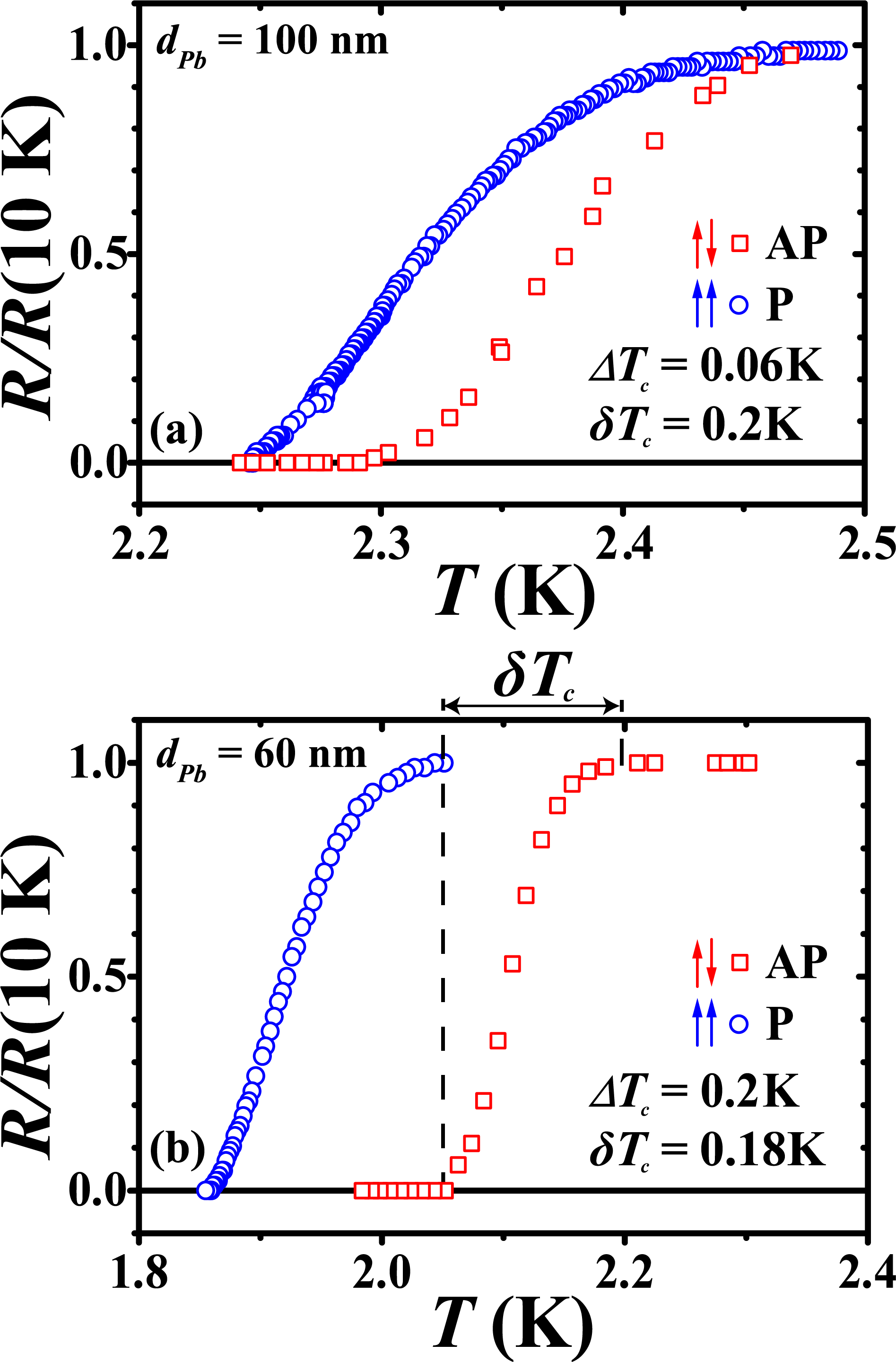}
\end{center}
\caption{Superconducting transition curves for the samples Pb\underline{ }100 (a) and Pb\underline{ }60 (b) at P ($H_0 = +1$\,kOe) ($\bigcirc$) and AP ($H_0 = -1$\,kOe) ($\square$)
 orientations of the Co1 and Co2 layers’s magnetizations, respectively.}
\label{fig2}
\end{figure}

Fig.~\ref{fig2} depicts the superconducting transition curves for the samples Pb\underline{ }100 (a) and Pb\underline{ }60 (b) at P ($H_0 = +1$\,kOe) and AP ($H_0 = -1$\,kOe) orientations of the Co1 and
Co2 layers' magnetization, respectively. The magnitude of the SSV effect for the sample Pb\underline{ }100  amounts to $\Delta T_c = 0.07$\,K whereas for the sample Pb\underline{ }60  it rises up to $\Delta T_c = 0.2$\,K. Obviously, the sample Pb\underline{ }60 demonstrates the full SSV effect since, as is evident from Fig.~\ref{fig2}(b), in this case $\Delta T_c > \delta T_c$.

\begin{figure}
\begin{center}
\includegraphics[width=0.5\columnwidth]{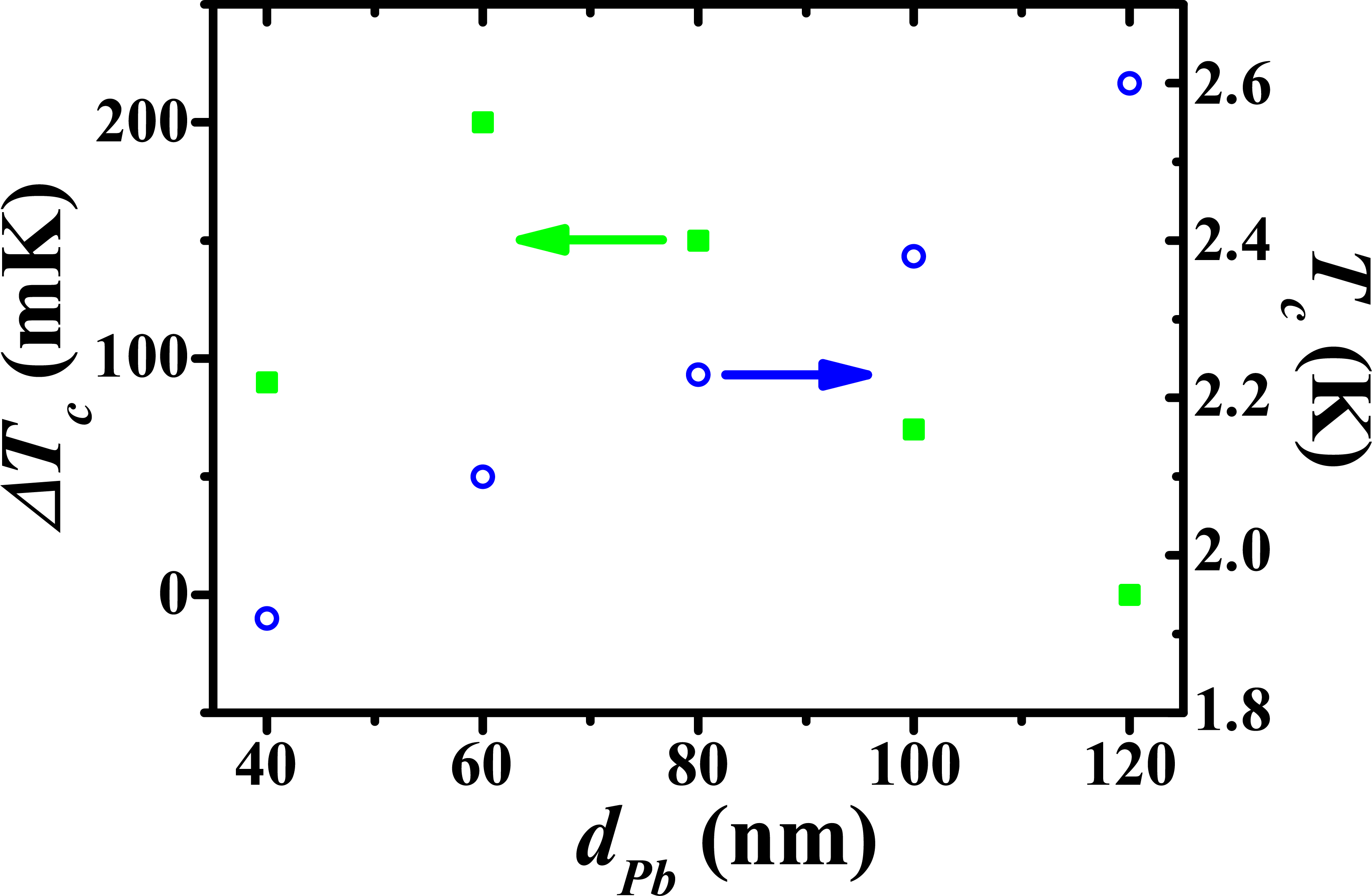}
\end{center}
\caption{The dependence of the magnitude of the SSV effect $\Delta T_c$ (left vertical scale) and of the superconducting critical temperature $T_c^\mathrm{P}$ for the parallel orientation of the magnetizations of the Co1 and Co2 layers (right vertical scale)  on the Pb layer thickness $d_\mathrm{Pb}$. }
\label{fig3}
\end{figure}

Fig.~\ref{fig3} shows the dependence of  $\Delta T_c$ and of $T_c$ on the thickness of the Pb layer $d_\mathrm{Pb}$ for the whole set of samples with insulating interlayers. $\Delta T_c$ increases and $T_c$ decreases approximately linearly with decreasing $d_\mathrm{Pb}$.
The maximum magnitude of the SSV effect $\Delta T_c = 0.2$\,K is reached at the minimum thickness of the superconducting layer $d_\mathrm{Pb}=60$\,nm.

Notably, in the control set of the structures with similar parameters, but without the insulating interlayers, superconductivity was not observed down to the lowest temperature of the experimental setup of
1.4\,K for all Pb thicknesses.

\section{Discussion}

\subsection{Phenomenology}

Our results demonstrate a significant SSV effect in the heterostructures  with insulating interlayers at the F1/S/F2 interfaces and finally verify the earlier observation by Deutscher and Meunier \cite{Deutscher}.
The S-layer thickness appears to be an essential parameter for observing the full SSV effect. As $d_\mathrm{Pb}$ decreases, the value of $\Delta T_c$ increases  and reaches its maximum of 0.2\,K at $d_\mathrm{Pb} =60$\,nm (Fig.~\ref{fig3}). The decrease of $T_c$ as a function of $d_\mathrm{Pb}$ is approximately linear down to $d_\mathrm{Pb} = 40$\,nm. A sharp drop of $T_c$ is expected at smaller thicknesses of the S layer due to the size effects (see, e.g., \cite{Kamashev2017}).
Apparently, the inverse
S/F proximity effect becomes  more pronounced as $d_\mathrm{Pb}$ decreases, despite the existence of insulating interlayers.
The here obtained value of $\Delta T_c$ at the optimal thickness of the Pb layer is twice as high compared to those found before in Refs.\ \cite{Leksin2012,Leksin2015,Garifullin,JMMM} for the structures with elemental metallic ferromagnetic layers but without insulating interlayers.

This observation is not trivial as it apparently contradicts the paramount prerequisite of the S/F proximity effect of having a perfect metallic contact between the S and F layers. It is plausible that oxide insulating  interlayers remain magnetic as suggested in Refs.\ \cite{Lommel1968,Deutscher}. They may play a dual role of attenuating the influence of the metallic ferromagnetic layer on the S layer that completely suppresses superconductivity in our control fully metallic Co1/Pb/Co2 stacks and at the same time maintaining some kind of the proximity effect that enables switching between the normal and superconducting states.

At the same time, the nontrivial S/F proximity effect in our system originates from metallic F layers. This is evidenced by the fact that similar control structures with reduced thickness of the Co layers (2\,nm instead of 3\,nm) did not show any spin-valve effect.


Note that the switching on/off superconductivity in the trilayer EuS/Al/EuS where EuS is a ferromagnetic insulator has been demonstrated by Li {\it et al.} \cite{Li2013}.
This type of system is different from metallic-type structures of Ref.\ \cite{Deutscher} and the present work, in which only very thin oxidized interfaces are insulating.

\begin{figure}
\begin{center}
 \includegraphics[width=0.5\columnwidth]{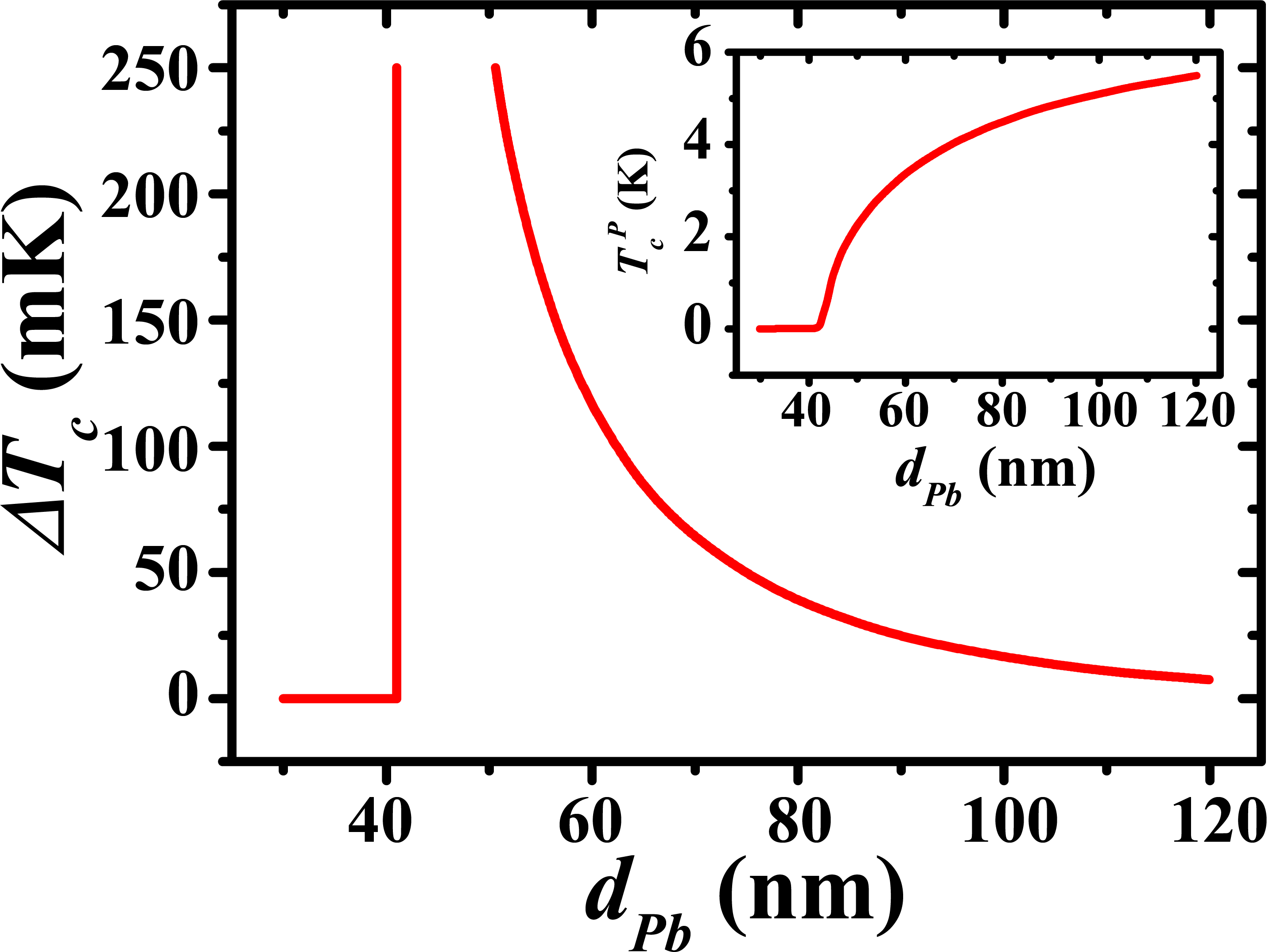}
\end{center}
\caption{Modeling results for  the $\Delta T_c(d_\mathrm{Pb})$ and $T_c^\mathrm{P}(d_\mathrm{Pb})$ dependences, main figure and inset, respectively.
The parameters of the model are as follows: $\xi_\mathrm{S} = 41$\,nm, $\xi_\mathrm{F} = 12$\,nm, $h = 0.035$\,eV, $\gamma = 0.093$, and $\gamma_b = 0.48$ (see the text and Ref.\ \cite{Fominov2003} for the exact definitions). }
\label{fig4}
\end{figure}

\subsection{Theoretical analysis}

The modeling of the observed significant SSV effect in heterostructures with intentionally deteriorated S/F interfaces is rather challenging due to an increased complexity of these interfaces as compared to the ideal metallic contacts between the layers. Though experimentally the formation of an insulating interlayer by oxidation appears to be a doable task, their characteristics, such as thickness, exact composition and physical properties, cannot be sufficiently well controlled at present. Therefore, in the following we will discuss if the main tendencies of the SSV effect with regard to the thickness of the superconducting Pb layer (Fig.~\ref{fig3}) could be at least qualitatively captured by theory.

The proximity-effect theory suitable for the description of $T_c$ in {\it symmetric} F1/S/F2 structures was formulated in Ref.\ \cite{Fominov2003}.
Applying this theory to our experimental data, under the above-mentioned somewhat ambiguous conditions we still
can achieve qualitative agreement with experiment, which captures the main tendencies demonstrated by the data.
In the case of the $\Delta T_c(d_\mathrm{Pb})$ dependence, see Fig.~\ref{fig4}, the theory demonstrates the nonmonotonicity of the dependence and approximate position of the maximum. This maximum is expected since the spin-valve effect should be suppressed both in the limit of very thin and very thick S layer, with a maximal value at some thickness $d_S$ of the order of the coherence length $\xi_S$. At the same time, the quantitative agreement between theory and experiment is not good, as expected. The same is true in the case of the $T_c^\mathrm{P}(d_\mathrm{Pb})$ dependence plotted in the inset to Fig.~\ref{fig4}.
The theoretical model predicts a sharp decline in $T_c$ at a $d_\mathrm{Pb}$ value close to 40 nm, but the measurements suggest a smoother dependence of the critical temperature possibly extending to lower temperatures. The model also suggests an asymmetric peak in $\Delta T_c(d_\mathrm{Pb})$, whereas a more symmetric peak is observed in Fig.~\ref{fig3} (at the same time, the left side of the peak is steeper than the right one both in theory and in experiment). Finally, the model indicates a rapid convex-type decay of $\Delta T_c(d_\mathrm{Pb})$ to the right of the peak, whereas the experimental data suggests a linear dependence. At the same time, due to limited number of experimental points, we cannot exclude a nonlinear behavior around the peak in $\Delta T_c(d_\mathrm{Pb})$ (in order to check this, points in the thickness range between 40 and 60 nm would be required).

The fitting parameters given in the caption to  Fig.~\ref{fig4} were obtained as follows. The coherence lengths in the S and F materials, $\xi_\mathrm{S}$ and $\xi_\mathrm{F}$, we estimated from the residual resistivities of the materials. The values of the exchange energy $h$, the materials-matching interface parameter $\gamma$, and the interface
resistance
parameter $\gamma_b$ were chosen in order to provide correct position of the $\Delta T_c$ maximum and acceptable overall values of this quantity (cf. Fig.~\ref{fig3}). The same values were then employed to plot the theoretical curve for $T_c^\mathrm{P}(d_\mathrm{Pb})$. The value of $\gamma$ is consistent with the values of $\xi_\mathrm{S}$ and $\xi_\mathrm{F}$.

What is unexpected in the above fitting parameters, is a rather small value of the interface
resistance parameter $\gamma_b= 0.48$.
In the tunneling limit, one can estimate $\gamma_b \sim t_b^{-1} l_\mathrm{F}/\xi_\mathrm{F}$ in terms of the effective interface transparency $t_b\ll 1$ (while $l_\mathrm{F}$ is the mean free path in the F material).
The transparency values are not directly measurable. At the same time, it is known that in the case of
conventional tunnel junctions with insulating interfaces of thickness
from 10 to 30 atomic layers, the order of magnitude of $t_b$ varies between $10^{-3}$ and $10^{-5}$ \cite{LikharevBook}. In that case, we would expect
larger values of $\gamma_b$ than the one resulting from our fitting. However, we have checked that larger $\gamma_b$ notably suppresses sensitivity of $T_c$ to the presence of the F layers (the inverse proximity effect becomes strongly suppressed).
The obtained small value of $\gamma_b$ points at small thickness of the tunneling barriers in our junctions.
Note that this correlates with observations by Deutscher and Meunier \cite{Deutscher}, who concluded that according to the resistance measurements, the barriers in their experiment were ``much thinner than in a conventional tunneling junction''.

While the theory \cite{Fominov2003} assumes symmetric F1/S/F2 structure, our samples may actually be asymmetric from the point of view of the interface transparencies. The oxidation times of the two interfaces were different in our samples and our fabrication procedure was such that the oxidation affected different materials (first, Co1 was oxidized, then Pb) at different temperatures. However, generalization of our theory to the case of two different $\gamma_b$ parameters is still expected to suppress the proximity effect almost completely in the case of two tunneling interfaces (while already one tunneling interface with $t_b \ll 1$ should effectively ``detach'' the corresponding F layer and thus suppress the effect of rotating magnetization on $T_c$).

A possible reason for not too small transparencies following from the fitting is that the insulating layers in our samples are actually very thin (a few atomic layers).
Another possibility is that the resulting oxides are not good insulators but possess finite conductivity or that metallic shortcuts are present inside the insulating layers.
Finally, in contrast to our theory \cite{Fominov2003} assuming nonmagnetic insulating barriers, the interfaces could be magnetically active \cite{Cottet2009}, which would introduce additional degrees of freedom into the problem (in particular, the interfaces could then nontrivially behave under the action of rotating magnetic field). Further experiments with better control of the insulating interfaces are clearly needed in order to clarify the role of the oxidized interfaces.

\section{Conclusions}

In summary, we have investigated superconducting properties of the Co1/Pb/Co2 SSV heterostructures with thin oxide insulating  interlayers formed at the  Co1/Pb and Pb/Co2 interfaces. We found the optimal thickness of the superconducting Pb layer for the realization of the full superconducting spin valve effect with the magnitude $\Delta T_c = 0.2$\,K. Our finding finally  substantiates the results of the earlier work by Deutscher and Meunier \cite{Deutscher} where a surprisingly large SSV effect was found for the F1/S/F2 structures with insulating interlayers.
It is remarkable that the here obtained value of $\Delta T_c$ significantly exceeds those of many of the multilayers prepared of elemental metallic ferromagnets and superconductors where special care was taken to achieve a perfect metallic contact at the S/F interface in order to enhance the S/F proximity effect.

At the same time, for the spin valve effect, the key role is played not by the strength of proximity effect but rather by {\it sensitivity} of the system to the variation of the relative magnetizations.
Our strategy was to achieve a ``fragile'' superconductivity which is sensitive to this kind of control. To this end, we have realized systems with such parameters that superconductivity is completely suppressed in the limit of perfect metallic interfaces. The role of the insulating interface layers is then to restore superconductivity in the system. This fragile ``restored'' superconductivity turns out to be indeed very sensitive to the configuration of the F part of the structure.

Our findings thus call for further exploration of this promising route to improve the operational parameters of the superconducting spin valves by advancing the preparation technologies and developing of underlying theories.


%
%
%

\bibliography{arXiv_bel}

\end{document}